\documentclass[prl,twocolumn,showpacs,superscriptaddress,amsmath]{revtex4}
\usepackage{graphicx}
\usepackage{amssymb}
\usepackage{citesort}

\begin{document}
\title{Theory of carrier concentration-dependent electronic behavior in layered cobaltates}
\author{H. Li}
\affiliation{Department of Physics, University of Arizona, Tucson, AZ 85721}
\author{R. T. Clay }
\affiliation{Department of Physics and Astronomy and HPC Center for Computational Sciences, Mississippi State University, Mississippi State, MS 39762}
\author{S. Mazumdar}
\affiliation{Department of Physics, University of Arizona, Tucson, AZ 85721}
\date{\today}
\begin{abstract}
A natural explanation for the carrier concentration-dependent
electronic behavior in the layered cobaltates emerges within
correlated-electron Hamiltonians with finite on-site and significant
nearest neighbor hole-hole Coulomb repulsions. The nearest neighbor
repulsion decreases hole double-occupancy below hole density
$\frac{1}{3}$, but increases the same at higher hole densities.  Our
conclusion is valid for both single-band and three-band extended
Hubbard Hamiltonians, 
%%HT3/21 - modifying in spite of small carrier
and sheds light on 
concentration-dependent $e_g^\prime$ hole occupancy within the latter.
\end{abstract}

\maketitle 
Layered cobaltates -- anhydrous Na$_x$CoO$_2$, Li$_x$CoO$_2$ and the “misfit”
cobaltates [Bi$_2$A$_2$O$_4$] · [CoO$_2$]$_m$, where A = Ba, Sr or Ca
and $m$ is incommensurate -- have attracted wide attention for their
unconventional metallicity and tunability of the carrier
concentration.  Na$_x$CoO$_2$ consists of edge-sharing CoO$_6$
octahedra, with CoO$_2$ layers separated by Na layers.  The Co ions
with average charge (4-$x$)+ form a triangular lattice.  Both
experiments \cite{Wang03a} and theory \cite{Singh00a} indicate large
crystal-field splitting and therefore low-spin states for the Co-ions.
Trigonal distortion splits the $t_{2g}$ $d$-orbitals into two
low-lying $e^{\prime}_g$ orbitals and a higher $a_{1g}$
orbital. Charge carriers are $S=\frac{1}{2}$ holes on the Co$^{4+}$
ion sites \cite{Hasan04a}.  The hole density $\rho = 1-x$ in anhydrous
Na$_x$CoO$_2$ and Li$_x$CoO$_2$.  Angle-resolved photoemission from
Na$_x$CoO$_2$ indicate that carriers occupy the $a_{1g}$ orbitals only
\cite{Hasan04a,Yang05a,Qian06a}, although Compton scattering finds
small $x$-dependent $e^{\prime}_g$ contribution \cite{Laverock07a}.

The electronic and magnetic behavior of these materials exhibit a
peculiar carrier concentration-dependence.  The temperature-dependent
magnetic susceptibility $\chi(T)$ in Na$_x$CoO$_2$ was early on
characterized as ``Pauli paramagnetic'' for $x<0.5$ and
``Curie-Weiss'' for $x>0.5$ \cite{Foo04a}.  The Curie-Weiss behavior
reflects strong Coulomb repulsion between the holes
\cite{Wang03a,Hasan04a,Yang05a,Foo04a,Vaulx07a}.  Strong correlation
at large $x$ is supported by observations of charge-ordering (CO)
\cite{Mukhamedshin05a, Lang08a}, Na-ion ordering \cite{Roger07a},
spin-density wave and intralayer ferromagnetic correlations
\cite{Bayrakci05a}.  Qualitatively different behavior for the small 
$x$ region is also agreed upon. Recent experimental work suggest that
(a) $\chi(T)$ here is weakly antiferromagnetic rather than Pauli
paramagnetic, and (b) the cross-over between strong and weak
correlations occurs at $x \sim 0.63 - 0.65$ rather than $x=0.5$
\cite{Lang08a}.  CO and Na-ion ordering are both absent for small $x$
\cite{Lang08a}.

Understanding the $x$-dependence of the electronic and magnetic
behavior of Na$_x$CoO$_2$ continues to be a theoretical challenge.  It
has been suggested that while the $a_{1g}$-only description is valid
for large $x$, holes occupy both $a_{1g}$ and $e^{\prime}_g$ orbitals
at small $x$ \cite{Lee05a}, and the on-site Hubbard correlation $U$ is
$x$-dependent.  Quantum chemical configuration interaction
calculations \cite{Landron10a} find larger $a_{1g}-e^{\prime}_g$
separation than LDA calculations \cite{Singh00a}, and (ii) many-body
approaches that take Coulomb hole-hole repulsion into account
\cite{Zhou05a,Marianetti07b,Bourgeois09a} do not find the
$e^{\prime}_g$ pockets on the Fermi surface predicted within LDA
calculations.  $x$-dependence has also been ascribed to differences in
the potential due to Na layers
\cite{Roger07a,Marianetti07a}. Experimentally, Li$_x$CoO$_2$
\cite{Motohashi09a} and Bi ``misfit'' cobaltates \cite{Bobroff06a} do
not exhibit ion ordering but nevertheless exhibit very similar carrier
concentration-dependence \cite{Motohashi09a,Bobroff06a}, suggesting
that this dependence is {\it intrinsic to the CoO$_2$ layers,} with
the interlayer potential playing a secondary role.

In the present Letter we show that a simple and natural explanation of
the carrier concentration-dependence emerges within $a_{1g}$-only as
well as multiband extended Hubbard models.  As we will be interested
in all three families of layered cobaltates including the misfits,
henceforth our discussions involve $\rho$ which is well-defined for
all systems, instead of $x$.  Following \cite{Marianetti07b} we write
the three-band Hamiltonian as
\begin{align}
&H=-\sum_{\langle ij\rangle\alpha\beta\sigma}t_{\alpha\beta}c^\dagger_{i\alpha\sigma}
c_{j\beta\sigma} +  \sum_{i}\Delta(n_{ie_g^\prime}-n_{ia_{1g}}) \nonumber \\
&+\frac{1}{2}\sum_{i\alpha\beta\sigma\sigma^\prime}
U_{\alpha\beta}^{\sigma\sigma^\prime}n_{i\alpha\sigma}n_{i\beta\sigma^\prime} 
+ V \sum_{\langle ij \rangle\alpha\beta} n_{i\alpha} n_{j\beta}.
\label{ham1}
\end{align}
Here $\alpha$ and $\beta$ refer to the $a_{1g}$ and $e_g^\prime$
orbitals, $c^{\dagger}_{i\alpha\sigma}$ creates a hole of spin
$\sigma$ on orbital $\alpha$ on site $i$,
$n_{i\alpha\sigma}=c^{\dagger}_{i\alpha\sigma}c_{i\alpha\sigma}$ and
$n_{i\alpha}=\sum_{\sigma}n_{i\alpha\sigma}$. $t_{\alpha \beta}$ is
the nearest neighbor (n.n.) hopping, $\Delta$ the trigonal splitting,
$U\equiv U^{\sigma,-\sigma}_{\alpha\alpha}$ and $U^\prime\equiv
U^{\sigma\sigma^\prime}_{\alpha\beta}$ are the onsite intra- and
inter-orbital Coulomb interactions, and $V$ is the n.n. Coulomb
interaction.  As in \cite{Marianetti07b}, we have ignored the Hund's
rule coupling based on the very small hole occupation of $e_g^\prime$
orbitals (see below).  As both photoemission experiments
\cite{Hasan04a,Yang05a,Qian06a} and many-body theories
\cite{Landron10a,Zhou05a,Bourgeois09a} find negligible role of
$e_g^\prime$ orbitals, we discuss the one-band limit of Eq.~\ref{ham1}
first.  We show that the $V$ term is essential within the one-band
model for understanding the $\rho$ dependence of the susceptibility.
We then show that the same effect not only persists in the full
three-band model, but influences the $\rho$-dependence of the
$e_g^\prime$ orbital occupation as well.

{\it Single-band limit.}  Terms containing $U^\prime$ and $\Delta$ are
irrelevant and $U$, $V$, and $t_{\alpha \alpha}$ refer to $a_{1g}$
orbitals only. We write $t_{\alpha \alpha}=t$ and express $U$ and $V$
in units of $t$. For hole carriers $t>0$ \cite{Hasan04a}.  Existing
$a_{1g}$-only theories largely assume $U \gg t$ \cite{Baskaran03a} or
$V=0$
\cite{Zhou05a,Marianetti07a,Bourgeois09a,Baskaran03a,Merino09a}. The
few studies that have investigated the effects of finite $V$ on
triangular lattices are either for particular $\rho=0.5$
\cite{Choy07a} or $\frac{2}{3}$ \cite{Watanabe05a}, or use approximate
approaches \cite{meanfield} that do not capture the complex
$\rho$-dependence that is seen experimentally and that we find in our
exact solutions.  We consider here realistic finite $U$ and $V$ and
investigate all $\rho$. We do not assume that $U$ and $V$ are
$\rho$-dependent. Rather, we show that {\it $\rho$-dependent
  correlations emerge as solutions to Eq.~\ref{ham1}.}

Two different observations give the appropriate parameter range.  (i)
At $\rho=1$ Eq.~\ref{ham1} can be replaced by a $V=0$ Hubbard
Hamiltonian with an {\it effective} on-site repulsion
$U_{\rm{eff}}=U-V$. Within the $V=0$ Hubbard model for the triangular
lattice, transition to the Mott-Hubbard insulator occurs for
$U_{\rm{eff}}>U_c$, where $U_c$ $\simeq 5-10$.
\cite{Mizusaki06a}. Experimentally, $\rho=1$ CoO$_2$ is a poor metal
close to the Mott-Hubbard transition \cite{Vaulx07a}, indicating that
$U-V \leq 5-10$ for cobaltates. (ii) For $V>\frac{1}{3}U$,
$\rho=\frac{2}{3}$ would be charge-ordered with all sites doubly
occupied (Co$^{5+}$) and vacant (Co$^{3+}$), as shown in
Fig.~1(d). The absence of such CO indicates $V<\frac{1}{3}U$. Taking
(i) and (ii) together, we conclude that the likely parameter regime is
$6<U<14$, $1<V<4$. Our estimate of $V/U$ is close to that of Choy {\it
  et al.}  \cite{Choy07a}. Our estimate of $U/t$ is slightly smaller
\cite{Lee05a,Bourgeois09a}.

\begin{figure}
\resizebox{3.0in}{!}{\includegraphics{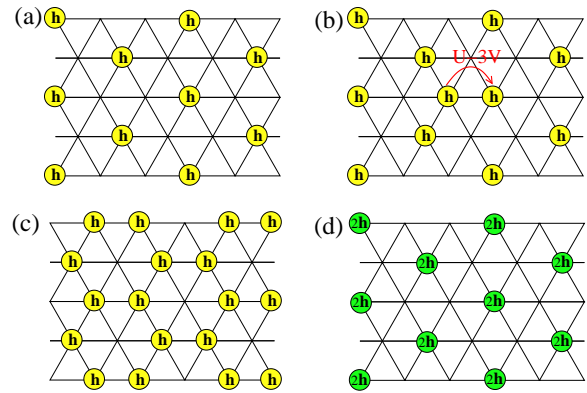}}
\caption{(Color online) Dominant ground state configurations for (a)
  $\rho=\frac{1}{3}$ and (b) one hole added to $\rho=\frac{1}{3}$.
Circles  labeled `h' (`2h') are singly (doubly) occupied sites, with
  vacancies occupying the vertices of the triangular lattice.  The
  n.n. hop indicated by the arrow in (b) costs
  $U-3V$. Dominant configurations for $\rho = \frac{2}{3}$, with (c)
  $N_d=0$, and (d) $N_d=\frac{1}{3}N$.} \label{config}
\end{figure}

We now argue that for realistic $U$ and $V$ there occur three distinct
hole density regions.  (i) $\rho \leq \frac{1}{3}$, where correlation
effects are strongest; (ii) intermediate $\frac{1}{3}<\rho \leq
\frac{2}{3}$, where correlation effects become weaker with increasing
density, and can be quite weak at the highest $\rho$; and (iii)
$\rho>\frac{2}{3}$, where correlation effects increase again slowly.
We classify configurations by the number of double occupancies $N_d$.
Fig.~\ref{config} (a) shows the $\sqrt{3} \times \sqrt{3}$ $N_d=0$
charge-ordered configuration that should dominate the ground state of
$\rho=\frac{1}{3}$. For fixed $U$, nonzero $V$ creates an energy
barrier to holes approaching each other.  The effective energy cost of
creating a double occupancy is thus {\it greater than $U$} for $\rho
\leq \frac{1}{3}$, which should exhibit strongly correlated behavior.
This situation changes as $\rho$ increases, as seen in
Fig.~\ref{config} (b), where we have added a single hole to the
charge-ordered configuration of $\rho=\frac{1}{3}$. The particular hop
indicated in the figure that creates a double occupancy costs only
$U-3V$. There are only three of these, and they increase $\langle N_d
\rangle$ very slightly.  With further increase in $\rho$, the number
of these low energy hops increases rapidly, increasing $\langle N_d
\rangle$.  In Fig.~\ref{config}(c) we show the two extreme
charge-ordered configurations for $\rho=\frac{2}{3}$, one with $N_d=0$
(Fig.~\ref{config}(c)), the other with $N_d=N_{max}=\frac{1}{3}N$. The
configurations are degenerate at $U=3V$.  {\it There is thus strong
  mixing of $N_d=0$ and $N_d>0$ configurations for $\rho$ close to
  $\frac{2}{3}$, even for $V<\frac{1}{3}U$.}  Adding double
occupancies to the configuration in Fig.~\ref{config}(d) is
prohibitively expensive in energy, which implies that for $\rho >
\frac{2}{3}$ the competition of $N_d=0$ is no longer with
$N_d=N_{max}$ but still with $N_d=\frac{1}{3}N$.  We expect
correlations to slowly increase again in this region.

We have performed exact numerical calculations to confirm the above
conjectures.  As a measure of correlations we have chosen the
normalized probability of double occupancy $g(\rho)$ in the ground
state,
\begin{equation}
g(\rho)=\frac{\langle n_{i,\uparrow}n_{i,\downarrow}\rangle}{\langle n_{i,\uparrow}\rangle \langle n_{i,\downarrow}\rangle}
\label{geq}
\end{equation}
g($\rho$)=1 and 0 for $U=0$ and $U \rightarrow \infty$, respectively,
for all $\rho$, and has intermediate values in between. $g(\rho)$ is
thus a measure of $U_{\rm{eff}}(\rho)$: small $g(\rho)$ implies
enhanced Curie-Weiss type $\chi(T)$ while moderate to large $g(\rho)$
implies weak antiferromagnetic spin-spin correlations
\cite{Mazumdar86a}.  Our proposed mechanism suggests that $g(\rho)$ is
small (large) for small (large) hole density, provided $V$ is
significant.

\begin{figure}
\resizebox{3.0in}{!}{\includegraphics{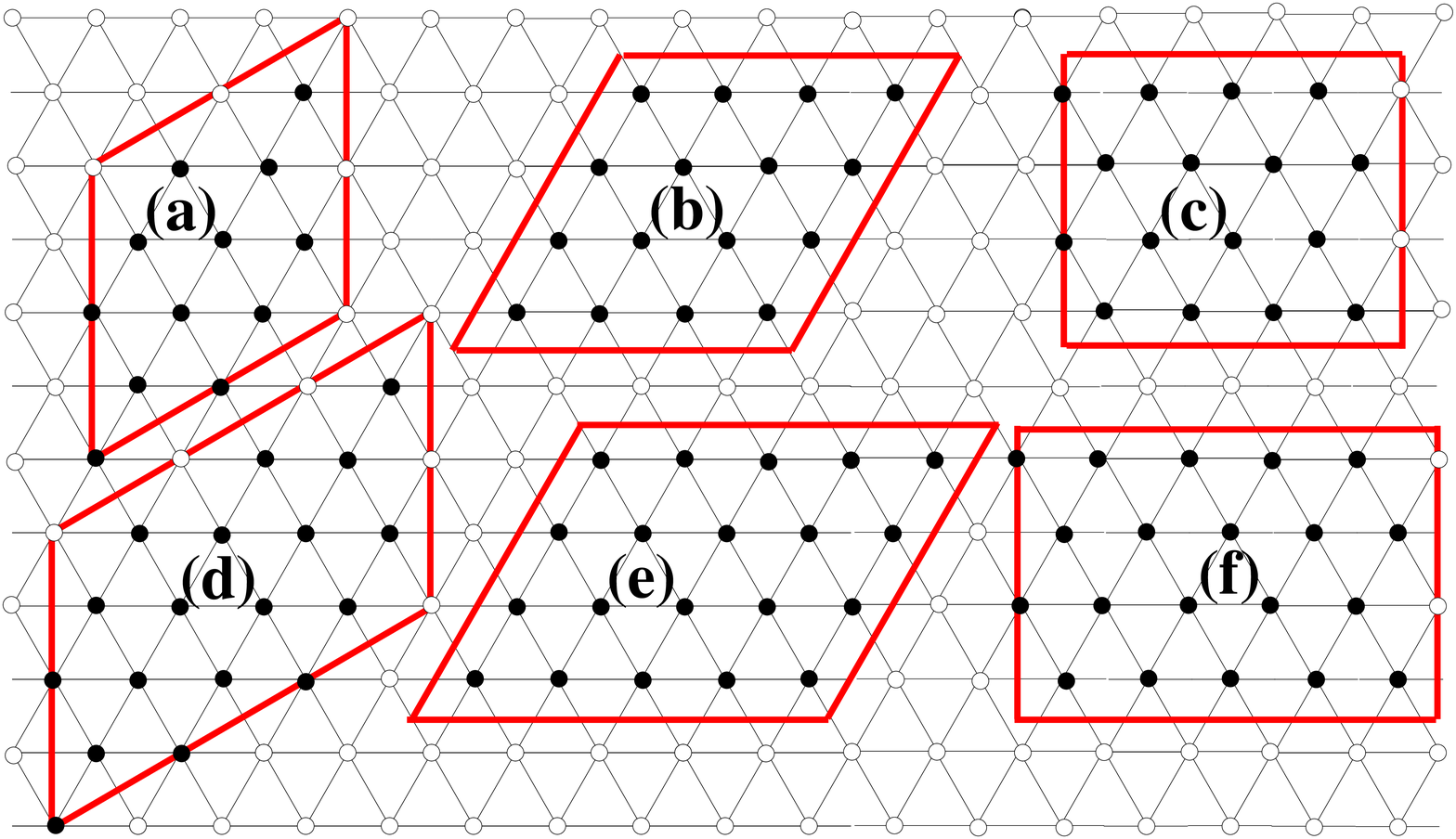}}
\caption{(Color online) 
Clusters investigated numerically: (a) $N=12$; (b) and (c) $N=16$;
(d) $N=18$; (e) and (f) $N=20$}
\label{lattices}
\end{figure}

We have calculated $g$ for the six triangular lattice clusters in
Fig.~\ref{lattices}, using periodic boundary condition.  For clusters
(a)-(c) the number of holes $N_{\rm h}$ covers the complete range
$\rho \leq 1$.  Computer memory constraints restrict us to $\rho \leq
0.88$ for cluster (d) and $\rho \leq 0.6$ for clusters (e) and (f),
respectively.  Our calculations are for all realistic $U$ and $V \leq
\frac{1}{3}U$.  As the results are qualitatively the same in all
cases, we report our results for $U=10$ only.

\begin{figure}
\resizebox{3.0in}{!}{\includegraphics{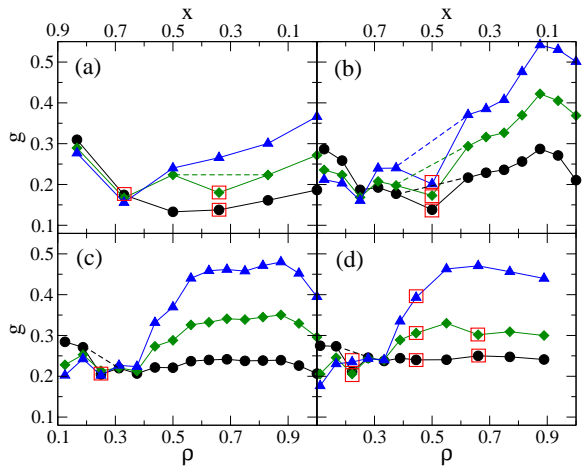}}
\caption{(Color online) Normalized probability of double occupancy of
  sites by holes versus hole density for $U$=10, $V$=0 (circles), 
2 (diamonds) and 3 (triangles) on clusters with
  (a) $N=12$, (b) and (c) $N=16$, corresponding to
  Fig.~\ref{lattices}(b) and (c), respectively, and (d) $N=18$. Points
within squares indicate $S>S_{\rm{min}}$ (see text).}
\label{g}
\end{figure}

In Fig.~\ref{g} we show our results for clusters (a)-(d) for $U=10$
and $V=0$, 2 and 3.  Our data points include both even and odd $N_{\rm
  h}$, and except for $N_{\rm h} \geq 14$ in Fig.~\ref{g}(d) we have
determined the total spin $S$ in the ground state in each case. With
few exceptions, $S=S_{\rm{min}}$, with $S_{\rm{min}}=0$
($\frac{1}{2}$) for even (odd) $N_{\rm h}$. 
%%SM3/21 - adding and modifying from here
$S>S_{\rm{min}}$ is a finite-size effect, as for different
clusters this occurs at different densities.
%with $S>S_{\rm{min}}$, 
The $g$-values for the $S>S_{\rm{min}}$ points were calculated
correctly in accordance with Eq.~\ref{geq}.
% both because of the
%``continuous'' nature of our plots and because the removal of these
%points from the figures will change neither the plots nor our overall
%conclusion.
The dips in $g$ for higher $S$ are expected.
In every case we have included dashed straight
lines connecting the neighboring points on both sides. The 
$g$ for $S=S_{\rm{min}}$ at these points is likely bounded by the computed points and the dashed lines.
%%to here

\begin{figure}
\resizebox{3.0in}{!}{\includegraphics{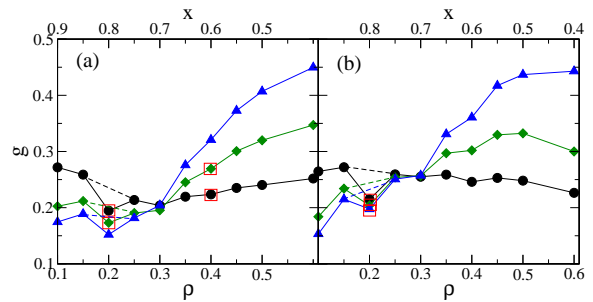}}
\caption{(Color online)
Same as Fig.~\ref{g} for $N=20$, for
$U=10$ and $V=0$ (circles), 2 (diamonds) and 3 (triangles); (a) and (b) correspond to 
lattices  Fig.~\ref{lattices}(e) and \ref{lattices}(f), respectively.}
\label{n20}
\end{figure}

In all four cases in Fig.~\ref{g}, $g(\rho)$ is nearly independent of
$\rho$ for $V=0$, but exhibits the $\rho$-dependence predicted for $V
\neq 0$.  The $\rho$-dependence is weakest for $N=12$.  
%%HT3/21 - modifying 
In all other cases there occur distinct strongly correlated low density region ($\rho \leq
0.4$), where $V$ suppresses $g$ and relatively weakly correlated intermediate density region
($0.4 \le \rho \le 0.8$), where $V$ enhances $g$.
%with {\it opposite effects of $V$ on
%  $g(\rho)$} are clearly visible.  
The predicted decrease in $g(\rho)$
for larger $\rho$ is also visible in Fig.~\ref{g}(b) and (c).

Fig.~\ref{n20} shows plots of $g(\rho)$ for the 20-site clusters of
Fig.~\ref{lattices}(e) and (f).  As in Fig.~\ref{g} we have retained
the points with $S>S_{\rm{min}}$.  {\it Distinct density regions (i)
and (ii), with opposite effects of $V$ are again clearly visible.}
The boundary between strongly and weakly correlated regions is $\rho
\approx 0.30$, in agreement with recent experiments \cite{Lang08a}.
Calculated charge-charge correlations $\langle n_in_j \rangle$ (not
shown) indicate that while near $\rho=\frac{1}{3}$ ($N_{\rm h}=6$ and
7) there is tendency to CO there is no such tendency at $\rho=0.5$
($N_{\rm h}=10$).  The qualitative agreement between our results for
six different clusters, {\it along with steeper $\rho$-dependence with
  increasing $N$,} strongly suggest that our results will persist in
the thermodynamic limit.

{\it Three-band model.} $\rho$-dependent $g$ is a consequence of the
competition between $U$ and $V$ at large $\rho$ \cite{Mazumdar86a} and
is unrelated to dimensionality or frustration.  Since exact
calculations are not possible for the clusters of Fig.~\ref{lattices}
within the three-band model, we have performed three-band calculations
for a one-dimensional (1D) periodic cluster with eight sites, each
with one ``$a_{1g}$'' and two ``$e_g^\prime$'' (hereafter $a$, $e_1$
and $e_2$, respectively) orbitals. The only differences between 1D and
the triangular lattice are, (i) the Wigner crystal occurs at
$\rho=\frac{1}{2}$ in 1D instead of $\rho=\frac{1}{3}$, and (ii) the
maximum in $g(\rho)$ is expected near $\rho=\frac{3}{4}$
\cite{Mazumdar86a} rather than $\frac{2}{3}$.  We retain the same $U$
and $V$ as for the single-band model.  We have taken
$t_{\alpha\alpha}=t$, and inter-orbital hopping
$t_{\alpha\beta}=0.1-0.3t$ ($\alpha\neq\beta$), $\Delta=3|t|$
\cite{Landron10a}, and $U^\prime=0.6U$ \cite{Bourgeois09a}.  As is
common for 1D rings, we use periodic (anti-periodic) boundary
conditions for $N_{\rm{h}}=4n+2$ ($4n$) \cite{suppl}.

\begin{figure}
\resizebox{3.0in}{!}{\includegraphics{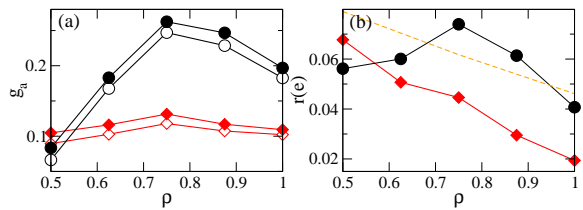}}
\caption{(Color online) (a) Normalized probability of double occupancy
  of the $a$-orbitals by holes in the 1D 3-band model for $N=8$,
  $\Delta=3$ (b) The fraction of holes occupying $e$-orbitals.  In
  both, unfilled (filled) symbols are for inter-orbital hopping 0.1$t$
  (0.3$t$). Diamonds (circles) are for $V=0$ ($V=3$).  In (b), the
  dashed line shows the noninteracting ($U=U^\prime=V=0$) result. For
  all other points $U=10$ and $U^\prime=6$.}
\label{multiband}
\end{figure}

In Fig.~\ref{multiband}(a), we have plotted $g_a(\rho)$, the
normalized probability of double occupancy of the $a$-orbitals by
holes, within the three-band model ($g_e(\rho)$ varies by less than
15\% over the entire range of $\rho$).  The carrier density $\rho$ here is
the ratio of the total number of holes and the number of $a$-orbitals,
in agreement with the definition of $\rho$ in Na$_x$CoO$_2$.  The
$g_a(\rho)$-behavior is nearly identical to that of $g(\rho)$ in
Figs.~\ref{g} and ~\ref{n20}.  Interestingly, $g_a(\rho)$ behavior is
the same for small and large $t_{e,a}$, {\it in spite of moderate hole
  population $n(e)$ in the $e$ orbitals in the latter case.}
Calculations for smaller $\Delta$ (not shown) indicate similar weak
dependence of $g_a(\rho)$ on $\Delta$.

In Fig.~\ref{multiband}(b) we plot $r(e)$, the fraction of holes that
occupy the $e$ orbitals, assuming $t_{\alpha\beta}=0.3t$, for
(a) noninteracting, (b) $U,U^\prime \neq 0$ but $V=0$, and (c) $V>0$ cases. For
$t_{\alpha\beta}=0.1t$, $r(e)\alt0.01$ and is negligible.  Comparing
the noninteracting and the $V=0$ plots, it is clear that nonzero $U$
and $U^\prime$ decrease the $e_g^\prime$ occupation, the reason for
which can only be correlation-induced band narrowing
\cite{Zhou05a,Bourgeois09a}. It is, however, the $V>0$ plot that is
far more interesting: $r(e)$ now shows a peak {\it at the same $\rho$
where $g_a(\rho)$ has a maximum}. This shows that correlation
effects due to $V$ play a much more complex role in the multiband
picture: in the large-$\rho$ region $V$ reduces $U_{\rm{eff}}$
(Fig.~\ref{multiband}(a)) for the same reason as in the single-band
picture. This reduces the extent of band-narrowing at precisely these
$\rho$, and leads to an {\it increase} in $r(e)$ {\it even for
$\rho$-independent $\Delta$.} Existing discussions of $e^\prime_g$
hole occupancy have largely focused on 
%experiments probing 
the Fermi surface \cite{Hasan04a,Yang05a,Qian06a,Laverock07a}.  
%%SM3/21 modifying from here 
%Our work is
%within a total-energy picture and hence difficult to directly compare
%to such experiments. Nevertheless, 
Our work shows that there can occur weak $e^\prime_g$ hole occupation even if
only the $a_{1g}$-orbitals form the Fermi surface. Most interestingly, 
the $e^\prime_g$ occupation estimated by Compton scattering experiments shows a $\rho$-dependence
(see Table I in \cite{Laverock07a}) that is very similar to our
results for $V>0$ in Fig.~\ref{multiband}(b), viz., {\it increasing}
$e^\prime_g$-occupancy with increasing $\rho$. This result simply cannot be
explained with $V=0$, as seen in Fig.~\ref{multiband}(b). Our calculations show that $\rho$-dependent
$\chi(T)$ and $e_g^\prime$ occupation are manifestations of the same
many-body effect.

{\it Summary.}  Strongly correlated behavior for small hole densities
and relatively weakly correlated behavior for larger hole densities
are both expected for nonzero n.n. Coulomb interaction.  To the best
of our knowledge there exists no other satisfactory theoretical
explanation for the observed weakly correlated behavior nearer to the
Mott-Hubbard semiconducting hole density and strongly correlated
behavior farther away from this limit.  The strong tendency to CO at
$\rho$ exactly $\frac{1}{3}$ and the absence of this tendency at
$\rho=\frac{2}{3}$ are both understood. The potential due to Na-ions,
ignored in our work, will strengthen the CO even for incommensurate
fillings with $\rho \leq \frac{1}{3}$ \cite{Roger07a}.  We do not find
CO at $\rho=0.5$, although it is moderately correlated.  Observed CO
here \cite{Foo04a} is likely driven by the cooperative effects of V
and the Na-ion potential. Conversely, the absence of Na-ion ordering
for weakly correlated $x<0.5$ in Na$_x$CoO$_2$ further suggests that
the CO and Na-ion ordering are synergistic effects.

This work was supported by the Department of Energy grant DE-FG02-06ER46315.

\end{document}